\documentstyle[twocolumn,aps,prl,times,epsfig]{revtex}

\begin{document}

\newcommand{\bx}{{\bf x}}

\draft
\title{Efficiency of a stirred chemical reaction
 in  a closed vessel}
\author{Crist\'obal L\'opez$^1$, Davide Vergni$^{1,2}$
 \and Angelo Vulpiani$^{1,2}$
}
\address{
  $^1$ Dipartimento di Fisica,
Universit\`a di Roma `La Sapienza',  P.le A. Moro 2, I-00185, Roma,
Italy.\\
  $^2$ INFM UdR and CSM Roma `La Sapienza',  P.le A. Moro 2, I-00185, Roma,
Italy.
}

\date{\today}
\maketitle

\begin{abstract}
We perform a numerical study of the reaction efficiency in a closed
vessel. Starting with a little spot of product, we compute the time
needed to complete the reaction in the container following an
advection-reaction-diffusion process.
Inside the vessel it is present a cellular velocity field
that transports the reactants. 
If the size of the container is not very large compared with the
typical length of the velocity field one has a plateau of the reaction
time as a function of the strength of the velocity field, $U$.  
This plateau appears both in the stationary and in the time-dependent flow.  
A comparison of the results for the finite system with the infinite case
(for which the front speed, $v_f$, gives a simple
estimate of the reacting time) shows the dramatic effect of the
finite size.

\end{abstract}


Numerous physical, biological and chemical systems show the
propagation of a stable phase into an unstable one~\cite{Xin,Murray}. 
When this phenomenon takes place in a fluid, one generally speaks of
front propagation in advection-reaction-diffusion (ARD) systems. Under this
generic name one indicates many different processes, e.g., the
propagation of plankton populations in ocean currents~\cite{Abra}, the
transport of reacting pollutants in the atmosphere
(e.g. ozone)~\cite{Epst}, or the premixed combustion~\cite{Peters}.

In the last years much effort has been done to study the influence of
an advection field on the front dynamics. In particular, it is well
established that the front speed in a laminar or turbulent fluid is
enhanced with respect to the propagation in a medium at
rest~\cite{Const,Kisel}.  In the context of (premixed) combustion
processes the flame front area is proportional to the front speed and,
therefore, an increasing of the front area due to the fluid stirring
gives rise to an enhancement of the burning efficiency, that is, the
{\it system burns faster}.

It is important to note that most of the theoretical studies and, in
particular, the above results about the enhancement of burning
efficiency, have been shown for an infinite-size system (in the
propagation direction).  Moreover,  in order to introduce
well-defined mathematical quantities one is forced to work with
an infinite (or with periodic boundary conditions) system. This is the
case of the front speed, which is an asymptotic quantity well defined
only for an  infinite system.

However, from a practical point of view one usually has to treat cases
where the size of the domain is not much larger than the typical
length scale of the velocity field~\cite{Boff}. 
The spreading of organisms in a lake or in a small closed sea basin,
or the combustion of fuel in a machine motor are two clear examples
where this may happen and, therefore, non-asymptotic properties can be
very important~\cite{Rys,experimental}.

In this work we treat the case of an ARD process confined in a closed
area.  Beginning with a small quantity of material in the stable phase
(in the following called burnt material), we  numerically compute 
the time needed for a given percentage of the total area to be also
burnt (called in the following, the reacting or burning time).  
The velocity field is of cellular flow type, that is, formed
by circulating cells of fluid. Both, the stationary and the chaotic
time-dependent cellular flow will be considered.  Our main result,
obtained either for the time-independent and for the time-dependent flow, 
is that increasing the typical velocity of the field one has a saturation of
the burning rate.  This saturation happens when the advection time scale
is much faster than the reaction time scale. 
Also, we compare our results with the infinite-size
case, studying the
crossover from finite size systems to the
asymptotic regime.
We observe that the relevance of the system size is more
important than expected {\it a priori}.

Let us consider the simplest
non trivial  case described by a scalar field
$\theta(\bx ,t)$ which represents the concentration of reaction
products, such that $\theta $ is equal one in the space-time
coordinates where the reaction is over (the stable phase), 
and $\theta$ is zero where there is fresh material 
(the unstable phase). The dynamical evolution of this field is given by
\begin{equation}
\partial_t \theta  + {\bf v} \cdot \nabla
\theta = D \nabla^2 \theta + \frac{f(\theta)}{\tau },
\label{ard}
\end{equation}
where ${\bf v} (\bx ,t)$ is the two-dimensional velocity field, $D$ is
the diffusion coefficient, and $f(\theta)$ is the reaction term, where
$\tau$ is the time scale for the reaction activity.  
For the reaction term we use the Fisher-Kolmogorov-Petrovskii-Piskunov
(FKPP) nonlinearity~\cite{FKPP37}, $f(\theta )= \theta (1- \theta)$.
Concerning the velocity field, ${\bf v}$, we first adopt a simple stationary
incompressible two-dimensional flow defined by the stream function
\begin{equation}
\psi (x,y)= \frac{UL}{\pi} \sin({\pi x \over L}) \sin({\pi y \over L}),
\label{sf}
\end{equation}
being the parameter $U$ the maximum vertical velocity of the flow,
and $L$ the size of one cell.
For a study of the transport properties in the field (\ref{sf})
see ref.~\cite{Gollub}; the asymptotic behaviour of front propagation
is discussed in ref.~\cite{Abel}.
The equations of motion for a fluid element are given by 
\begin{equation}
   \left \{ {\begin{array}{lcr}
               \dot x & = & \partial_y \psi\\
               \dot y & = & -\partial_x \psi .
             \end{array} } \right .
   \label{eq:veloc} 
\end{equation}
In this work the reaction processes described by (\ref{ard}) take
place in a closed recipient.  This confinement is implemented by
assuming rigid boundary conditions on the box $0 \le y \le L$ and $0
\le x \le n L$, where $n$ is the number of circulating cells of the flow.
One  approaches  to the asymptotic case  increasing the
value of $n$.

The settling of the problem is completed when we indicate the initial
conditions, i.e., an initial spot of burnt material which
starts the reaction. Thus, we use in all our numerical
experiments a small circle of radius $r$ filled with stable material
($\theta = 1$), that is placed at the initial time in the box filled
with unstable material ($\theta =0$). The initial coordinates of the
center of this circle are $(x=r, y=L /2)$ (the circle is on the border
of the box; this mimic the injection of reacting material from the
outside).  
As anticipated, the principal observable under investigation
is the time needed for a given percentage of the total area to be
burnt. We define $S(t) =\frac{1}{\Delta S} \int_{\Delta S} 
{\mathrm d}x {\mathrm d}y \theta (x,y,t)$ 
as the percentage of area burnt at time $t$, where $\Delta S= L^2 n$
is the total area of the container. In our case, by choosing an
appropriate $r$ the initial burnt material is $S(0) = 0.005/n$, which
is the $0.5\%$ of one cell.  The reacting or burning time $t_\alpha$
is defined as the time needed for the percentage $\alpha$ of the total
area of the recipient to be burnt, i.e., $S(t_\alpha)=\alpha $.

Numerically, to integrate (\ref{ard}) we use the Feynman-Kac (FK) or
stochastic Lagrangian approach. In this algorithm the field
evolution is computed using the Lagrangian propagator plus a
Montecarlo integration for the diffusive term. Then, the reaction
propagator accounts for the reacting term (for details
see~\cite{Abel,Fmnv}). We also impose a rigid wall condition in the
boundaries, in order to avoid that any fluid particle leaves the
container, which could happen due to the noise term added
to the velocity field in the Lagrangian approach.

\begin{figure}
\epsfig{file=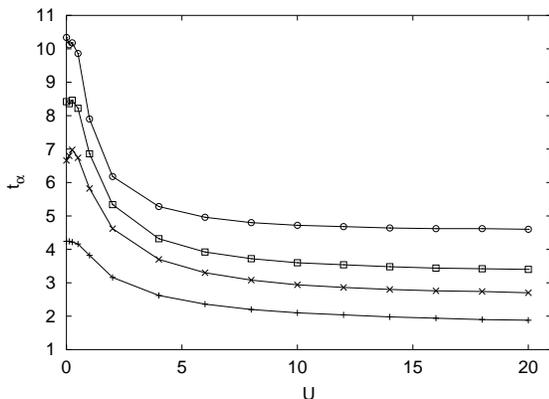,width=0.9\linewidth }
\caption{$t_{\alpha}$ against the flow strength $U$ with $\tau=0.4$,
$n=1$ and for various $\alpha$: $\alpha=0.2$ $(+)$, $\alpha=0.5$
$(\times)$, $\alpha=0.7$ $(\Box)$, $\alpha=0.9$ $(\circ)$.}
\label{plateau}
\end{figure}

We first show in Fig.~(\ref{plateau}) the influence of the velocity on
the reacting efficiency, when different percentages, $\alpha$, of the
final burnt area are considered.
Increasing the velocity of the flow, $U$,
$t_{\alpha}$ decreases monotonically until a plateau is reached.  
Then, a further increasing of the flow velocity ($U > 15$, similar for
different $\alpha's$) does not decrease the burning time $t_\alpha$.
We remark that this effect also appears for different finite values of the
system size (different $n$'s), and different chemical rates $\tau$.
At first glance, the appearance of the plateau seems to be surprising: 
{\it in a closed container and for very high stirring intensity,
increasing further more this intensity there is not an enhancement in
the burning front propagation, that is, one does not improve the
efficiency of burning}.

The existence of the plateau can be understood noting that it 
is reached only when the reaction time, $\tau$, is large compared with
the advection time, $\tau_a = L/U$.  In this case, in the first
stages of the process the active material invades the whole container
because of advection and diffusion.
Then the reaction term is the final responsible for the cell burning.

\begin{figure}
\epsfig{file=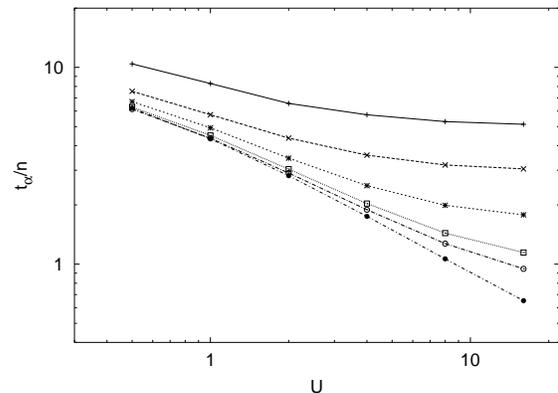,width=0.9\linewidth}
\caption{The burning time per unit-cell, $t_{\alpha}/n$,
at various $U$  for $\tau=0.4$ and 
$\alpha=0.9$. The plots are for different system's sizes: 
$n=1$ $(+)$, $n=2$ $(\times)$, $n=4$ $(*)$, $n=8$ $(\Box)$ and
$n=12$ ($\circ$). It is also shown ($\bullet$) the
burning time calculated using the front speed: $t=nL/v_f$.}
\label{sizefig}
\end{figure}

A direct comparison between the finite and the infinite system is
interesting. In Fig.~(\ref{sizefig})  we show
the burning time scaled with the system size, i.e.,
$t_{\alpha}/n$, against the typical flow velocity, $U$, for some values
of $n$.  We also we plot the data obtained for an infinite system,
which have been calculated from the front speed $v_f$ of the infinite
system data according to
\begin{equation}
   t_\alpha \simeq \frac{nL}{v_f},
   \label{eq:astime}
\end{equation}
which  is expected to hold for  $\alpha$ close to $1$ and large $n$.
Figure~(\ref{sizefig}) shows that the asymptotic reacting time (given
by $v_f$) is reached only in the large size limit, i.e., $n$ large,
while the dynamics of small systems is dominated by the non asymptotic
properties of the evolution. 

Since the considered bidimensional velocity field (\ref{sf}) is
stationary the Lagrangian trajectories are not chaotic. 
Nevertheless also in the case of Lagrangian chaotic trajectories,
obtained with a time-dependent flow, the burning time shows the same
qualitative behaviours shown in figures (\ref{plateau}) and
(\ref{sizefig}). Let us consider a time dependent flow whose
streamfunction is
\begin{equation}
\psi (x,y,t)= U \sin (x+ B(x) \cos(\omega t) ) \sin(y)\,.
\label{nssf}
\end{equation}
This is sufficient to induce Lagrangian chaos~\cite{LagChaos} in the evolution
of passive tracers advected by the velocity field (\ref{eq:veloc})
generated by (\ref{nssf}).  Because we are dealing with closed systems
$B(x)$ is constructed in such a way that it is zero near the boundary
of the system and almost constant, $B_0$, otherwise: $B(x) = B_0(1 -
\exp(-1/x) - \exp(-1/(nL - x)))$.
\begin{figure}
\epsfig{file=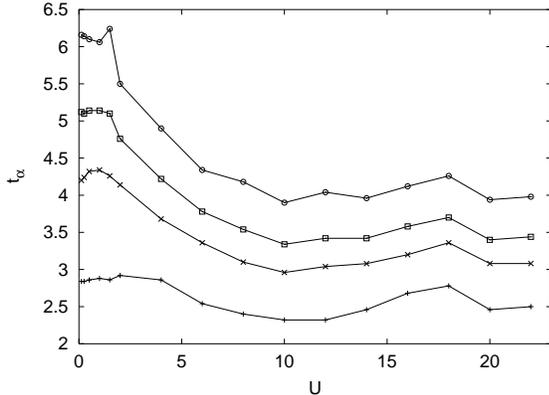,width=0.9\linewidth}
\caption{$t_{\alpha}$ against the flow strength $U$ for various
$\alpha$ in the unsteady case ($B=1.1$ and $\omega=2.09$): 
$\alpha=0.2$ $(+)$, $\alpha=0.5$ $(\times)$, $\alpha=0.7$ $(\Box)$, 
$\alpha=0.9$ $(\circ)$. The flow is confined in two cells.}
\label{fig:plateau_td}
\end{figure}
In Fig.~(\ref{fig:plateau_td}) we show the curve $t_\alpha$ against $U$
for different values of $\alpha$. At difference from the previous
case, when unsteady flow is concerned there is not a simple plateau
in $t_\alpha$, but an oscillatory behaviour due to the interplay
between the oscillation period of the separatrices and the circulation
time inside the cell. It happens that circulation and oscillation
``synchronize'' producing a very efficient and coherent way of
transferring passive particles from one cell to the other.
A similar, but much more impressive, feature occurs for the effective
diffusion coefficient in the horizontal direction~\cite{casti}.
Anyway, for low values of $U$ the mixing induced by the time
dependence makes that the system burns quite faster.  For higher $U$, the
mixing properties of the flow are not sufficient to improve
furthermore the burning efficiency, and at the end, the typical time
for the cell burning is proportional to the reaction time-scale.

Summing up, the physical mechanism of reaction in a closed recipient
for both time-independent and time-dependent cellular flows are quite
similar.  When the typical time-scale of the velocity field, $L/U$, is
larger than the reaction time-scale, $\tau$, the initial condition is
quickly spread along the whole system due to advection (and diffusion), 
in such a way that in all the container the value of $\theta$ is small
but different from zero.
Then, basically only reaction plays an important role, and a further
increasing of the flow velocity does not decrease the burning time.

Let us study the dependence of the saturation time
$t_{\alpha}^s(\tau)$, which is the value of the burning time in the
plateau, as a function of $\tau$. In the unsteady case, we choose the
minimum value of $t_{\alpha}$ at varying $U$ as the saturation time.

\begin{figure}
\epsfig{file=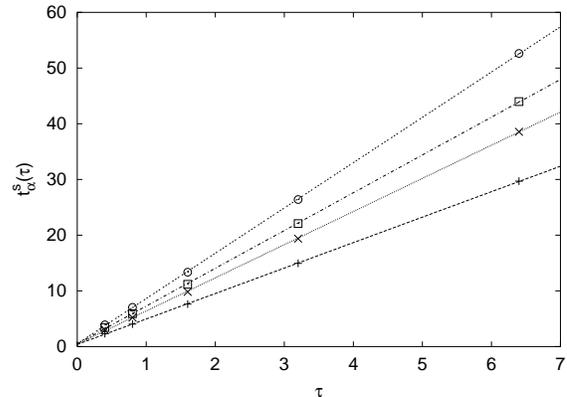,width=0.9\linewidth }
\caption{Plot of $t_{\alpha}^s (\tau)$ vs $\tau$ at various percentages
of filling $\alpha$ in the time-dependent case ($B=1.1$ and $\omega=2.09$):
$\alpha=0.2$ $(+)$, $\alpha=0.5$ $(\times)$, $\alpha=0.7$ $(\Box)$, 
$\alpha=0.9$ $(\circ)$. For each curve we have superimposed 
the linear fit $t_{\alpha} ^s(\tau) = \tilde t + b_\alpha \tau$, 
where $\tilde t$ and $b_\alpha$ are fitting parameters. Here $n=2$.}
\label{vartau}
\end{figure}

In Figure~(\ref{vartau}) we show the result for the unsteady case,
which can be interpreted following the same arguments used to explain
the existence of the plateau. Essentially the burning process
in the case of large $U$ can be divided in two steps, the initial
mixing regime and the reacting dominated regime.  In fact, from the
initial condition (in which only a small fraction of the volume is
active) one has a rapid spreading (say in a time $\tilde t$) because
of advection and diffusion.  After the spreading one has an
exponential increasing of the field due to the reaction term.  The
time $\tilde t$ is expected to be a decreasing function of $U$ (with a
limiting value for sufficiently large $U$).  This implies, together
with a dimensional argument, $t_{\alpha}^s (\tau) = \tilde t + b_{\alpha}
\tau$.  Thus, there is a linear dependence of $t_\alpha^s$ with $\tau$
as shown in Figure~(\ref{vartau}), which has been obtained for the
time-dependent flow, but similar results hold also for the steady case.
The value of the slope of the saturation time against $\tau$,
$b_\alpha$, can be analytically studied.  We have that, in the regime
of very high $U$ such that the plateau is reached, after the time
$\tilde t$ the initial condition is spread out through the whole
container and one can approximate $\theta (x,y,t) \sim \tilde \theta(t)$, 
being $\tilde \theta (t)$ a rough average of the $\theta$ field in the
container, which evolves following only the reaction part of (\ref{ard}). 
This is because, as we have argued before, after $\tilde t$ 
the important physical mechanism of burning comes from the chemical
activity and not from the mixing due to the advection and
diffusion. Then $S(t)=\tilde \theta (t)$ for $t>\tilde t$, and so:
\begin{equation}
\frac{{\mathrm d}S}{{\mathrm d}t}={1 \over \tau}S (1-S).
\label{nuevaS}
\end{equation}
This can be integrated from $\tilde t$ to $t$, taking into account that 
we can approximate $S(\tilde t) \sim S(0)$, i.e.,  the initial condition is
just spread out in the system in the initial stages of the process. 
One has that
\begin{equation}
\log \left( \frac{S(t)}{1-S(t)} \right)-
\log \left( \frac{S(0)}{1-S(0)} \right)=\frac{t-\tilde t}{\tau} \,\,.
\label{nueva2S}
\end{equation}
Finally, as $S(t_\alpha^s)=\alpha$ we get
\begin{equation}
t_\alpha^s = \tilde t + \tau \log \left( \frac{\alpha (1-S(0))}{(1-\alpha)S(0)} \right)
 \equiv \tilde t + b_\alpha \tau\,\,,
\label{slopeb}
\end{equation}
which gives the dependence of $b_\alpha$ on the percentage
of burnt material, $\alpha$.

For the time-independent cellular flow the reasonings follow 
closely the former ones. In fact, in the regime of large $U$,
when all over the cell (by diffusion) there is also a small quantity 
of active reagent, the reaction process can begin with 
an averaged $S(t)$ (in a mean-field sense).

\begin{figure}
\epsfig{file=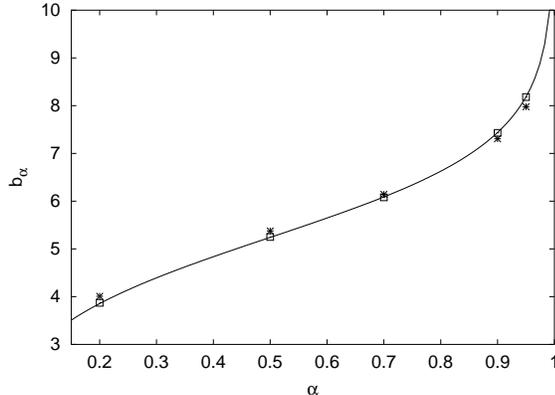,width=0.9\linewidth}
\caption{The slope, $b_\alpha$, of the saturation time
$t_{\alpha}^s$ against $\tau$, that is, the slope of the curves in
Fig.~(\ref{vartau}) (with an additional value for $\alpha=0.95$), vs the
percentage $\alpha$. With $(\Box)$ the numerical values for time
dependent flow, with $(*)$ the numerical value (rescaled) for
steady flow, and with the solid line the prediction given by (\ref{slopeb}).}
\label{fitcris}
\end{figure}

Despite the numerous approximations done to obtain (\ref{slopeb}), it
is in excellent agreement with the numerical results, see
Fig.~(\ref{fitcris}), confirming once again the physical mechanism we
think that give rise to the existence of the saturation time.
Summarizing, we have performed a numerical study of an
advection-reaction-diffusion system confined in a closed vessel, 
using stationary and time dependent cellular flows.
Beginning with a small quantity of the active phase, we have
calculated the time needed for a percentage of the total area to be
burnt. Thus, our numerical experiments may represent the spreading of
an organism in a lake or the combustion of a material in a vessel.
The main lesson to learn from our studies is that the influence of the
system size has been shown to be very important~\cite{fedotov}. In particular, we
have shown that for very large flow velocities the reacting time
saturates, giving rise to the unexpected result that increasing
further more the flow velocity there is not a decreasing of the time
to burn the material.  
We have to mention that a similar scenario, i.e., the appearance of
the plateau in the burning time, has been obtained for other types of
chemical reactions $f(\theta)$, like the Arrhenius $f(\theta )=
(1-\theta)\exp(-\theta_0/\theta)$ ($\theta_0$ constant) or the Zeldovich
function $f(\theta) = \theta^m (1-\theta)$, with $m=2$.

This work has been partially supported by
INFM {\it Parallel Computing Initiative and MURST (Cofinanziamento
Fisica Statistica e Teoria della Materia Condensata)}. 
C.L. acknowledges
support from MECD of Spain, D.V and A.V. acknowledge support
from the INFM Center for Statistical Mechanics and
Complexity (SMC).


\begin{thebibliography}{xx}

\bibitem {Xin} J. Xin, 
SIAM Review {\bf 42}, 161 (2000). 

\bibitem {Murray} J.D. Murray, {\it Mathematical Biology}, 
Springer-Verlag, Second Edition, (1993).
\bibitem{Abra} J.H. Steele (ed.),
{\it Spatial patterns in plankton communities},
Plenum Press, New York (1978).


\bibitem{Epst} I.~R.~Epstein, 
Nature {\bf 374}, 231 (1995).

\bibitem{Peters} N.~Peters, 
{\it Turbulent combustion},
Cambridge University Press (2000).

\bibitem{Const} P.~Constantin, A.~Kiselev, A.~Oberman and L.~Ryzhik,
Arch. Rational Mechanics {\bf 154}, 53 (2000).

\bibitem{Kisel} A.~Kiselev and L.~Ryzhik, 
Ann. I.H. Poincar\'e {\bf 18}, 309 (2001).

\bibitem{Boff} G.~Boffetta, A.~Celani, M.~Cencini, G.~Lacorata and A.~Vulpiani,
Chaos {\bf 10}, 50 (2000).

\bibitem{Rys} P.~Rys, 
Chimia {\bf 46}, 469 (1992).

\bibitem{experimental}
A. Gorczakowski, A. Zawadzki, J. Jarosinski and
B. Veyssiere,
Combust. Flame {\bf 120}, 359 (2000).

\bibitem{FKPP37} A.~N.~Kolmogorov, I.~G.~Petrovskii, and N.~S.~Piskunov, 
Moscow Univ. Bull. Math. {\bf 1}, 1 (1937); \\
R.~A.~Fischer, 
Ann. Eugenics {\bf 7}, 355 (1937).

\bibitem{Gollub} T.H.~Solomon and J.P.~Gollub, 
Phys. Rev. A {\bf 38}, 6280 (1988). 

\bibitem{Abel}  M.~Abel, A.~Celani, D.~Vergni and A.~Vulpiani, 
Phys. Rev. E {\bf 64}, 046307 (2001).

\bibitem{Fmnv}
U. Frisch, A. Mazzino, A. Noullez and M. Vergassola,
Phys. Fluids {\bf 11}, 2178 (1999).

\bibitem{fedotov}
M.~V.~Tretyakov and S.~Fedotov, Physica D {\bf 159}, 190 (2001).

\bibitem{LagChaos} H.~Aref, J.~Fluid Mech. {\bf 143}, 1 (1984);\\
A.~Crisanti, M.~Falcioni, G.~Paladin and A.~Vulpiani, 
Riv. Nuovo Cimento {\bf 14}, 1 (1991).

\bibitem{casti}
P. Castiglione, A. Crisanti, A. Mazzino, M. Vergassola and
A. Vulpiani, J. Phys. A {\bf 31}, 7197 (1998).

\bibitem{fedotov}
 A different phenomenon where the influence of the
system size is also very important is reported in 
M.~V.~Tretyakov and S.~Fedotov, Physica D {\bf 159}, 190 (2001).


\end{thebibliography}
\end{document}